\documentclass{iaus}
\usepackage{graphicx}

\title[Extragalactic Star Clusters] 
{Formation History of Stars and Star Clusters in Nearby Galaxies}

\author[S. S. Larsen et al.] 
{S. S. Larsen$^1$, M. D. Mora$^2$, J. P. Brodie$^3$ \and T. Richtler$^4$}   

\affiliation{$^1$Astronomical Institute, University of Utrecht, 
  Princetonplein 5, 3584 CC, Utrecht, The Netherlands,
  email: larsen@astro.uu.nl \\ [\affilskip]
$^2$European Southern Observatory, Karl-Schwarzschild-Strasse 2,
  85748 Garching bei M{\"u}nchen, Germany,
  email: mmora@eso.org \\ [\affilskip]
$^3$UCO/Lick Observatory, 1156 High Street, University of California,
  Santa Cruz, 95064 CA, USA,
  email: brodie@ucolick.org \\ [\affilskip]
$^4$Astronomy Group, Universidad de Concepci{\'o}n, Departamento de 
  F{\'i}sica, Casilla 160-C, Concepci{\'o}n, Chile,
  email: tom@mobydick.cfm.udec.cl
}

\pubyear{2006}
\volume{241}  
\pagerange{xxx--yyy}
\date{?? and in revised form ??}
\jname{Stellar Populations as Building Blocks of Galaxies}
\editors{A.\ Vazdekis \& R.\ Peletier}
\begin{document}

\maketitle

\begin{abstract}
We present first results from an HST/ACS imaging survey of stars and star 
clusters in five nearby spiral galaxies. This contribution concentrates
on NGC~1313, a highly distorted late-type barred spiral. We compare the
field star and cluster formation histories in
our three ACS pointings for this galaxy. In one pointing, both the
cluster and field star age distributions show clear evidence for a ramp-up in 
the star formation rate about $10^8$ years ago.
\end{abstract}

\firstsection 
\section{Introduction}

Star clusters offer a potentially powerful alternative to ``field'' stars 
for studies of star formation histories in external galaxies.
The most massive of them are composed of $10^5-10^6$ or more
stars, and can be observed at much greater distances than individual
stars. Globular clusters have survived since very early in the history
of the Universe, so clusters clearly have the 
potential to trace virtually the \emph{entire} 
evolutionary histories of galaxies. Furthermore, they offer the 
distinct advantage that their integrated spectra are typically broadened by 
only a few km/s, compared to hundreds of km/s for massive galaxies, 
so the prospects are bright for taking full advantage of the high-resolution
stellar and SSP model spectral libraries that are now becoming available.
Efforts are currently under way by several groups to develop techniques for 
detailed abundance analysis based on high-dispersion spectroscopy of 
extragalactic young and old star clusters (R.\ Peterson, these proceedings;
\cite[Bernstein \& McWilliam 2005; Larsen et al.\ 2006]{bw05,lar06}).

However, star clusters come with their own set of challenges. 
One big challenge is to explain the 
large variations in globular cluster \emph{specific frequency}, not just 
between galaxies but also for stellar populations within galaxies. This is
nicely illustrated by the Fornax dwarf spheroidal, 
whose field star metallicity distribution peaks at [Fe/H]$\sim-1.0$
\cite[(Helmi et al.\ 2006; Tolstoy, these proceedings)]{helmi06} while the 
5 globular clusters all are significantly more metal-poor ([Fe/H]$\sim-2.0$; 
\cite[Strader et al.\ 2003; Letarte et al. 2006]{stra03, let06}). The picture
is complicated by dynamical evolution, which tends to 
selectively remove low-mass clusters 
\cite[(Fall \& Zhang 2001; Larsen 2006)]{fz01,larsen06}.
Unless these differences are understood, it will be difficult to infer
the relative and absolute \emph{strengths} of star formation episodes from 
studies of cluster populations, although the \emph{timing} and age-metallicity
relations may be constrained with greater confidence.

\begin{figure}
  \centerline{\includegraphics[width=66mm]{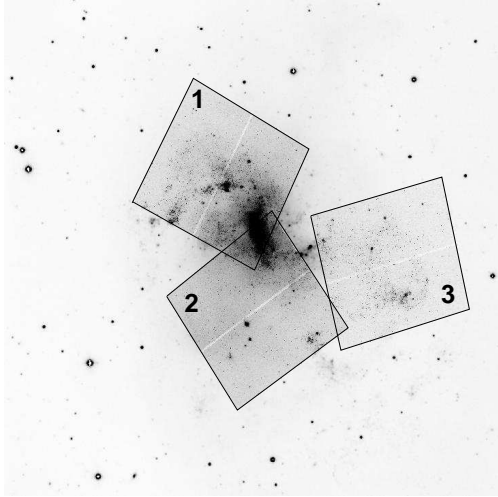}}
  \caption{\label{fig:n1313}NGC~1313 and our three HST/ACS pointings.}
\end{figure}

\begin{figure}
  \noindent \includegraphics[width=14cm,height=4.5cm]{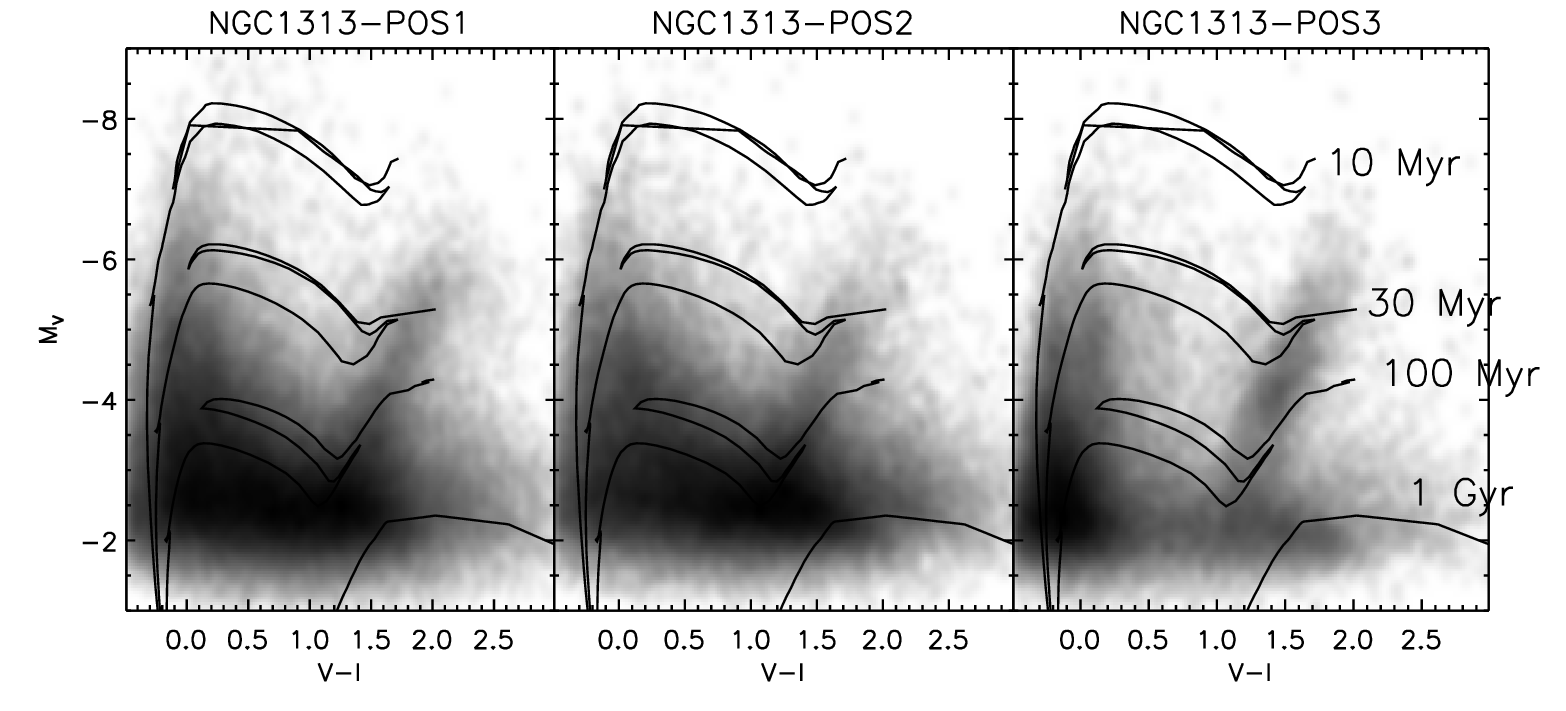} \\
  \noindent \includegraphics[width=14cm,height=4.5cm]{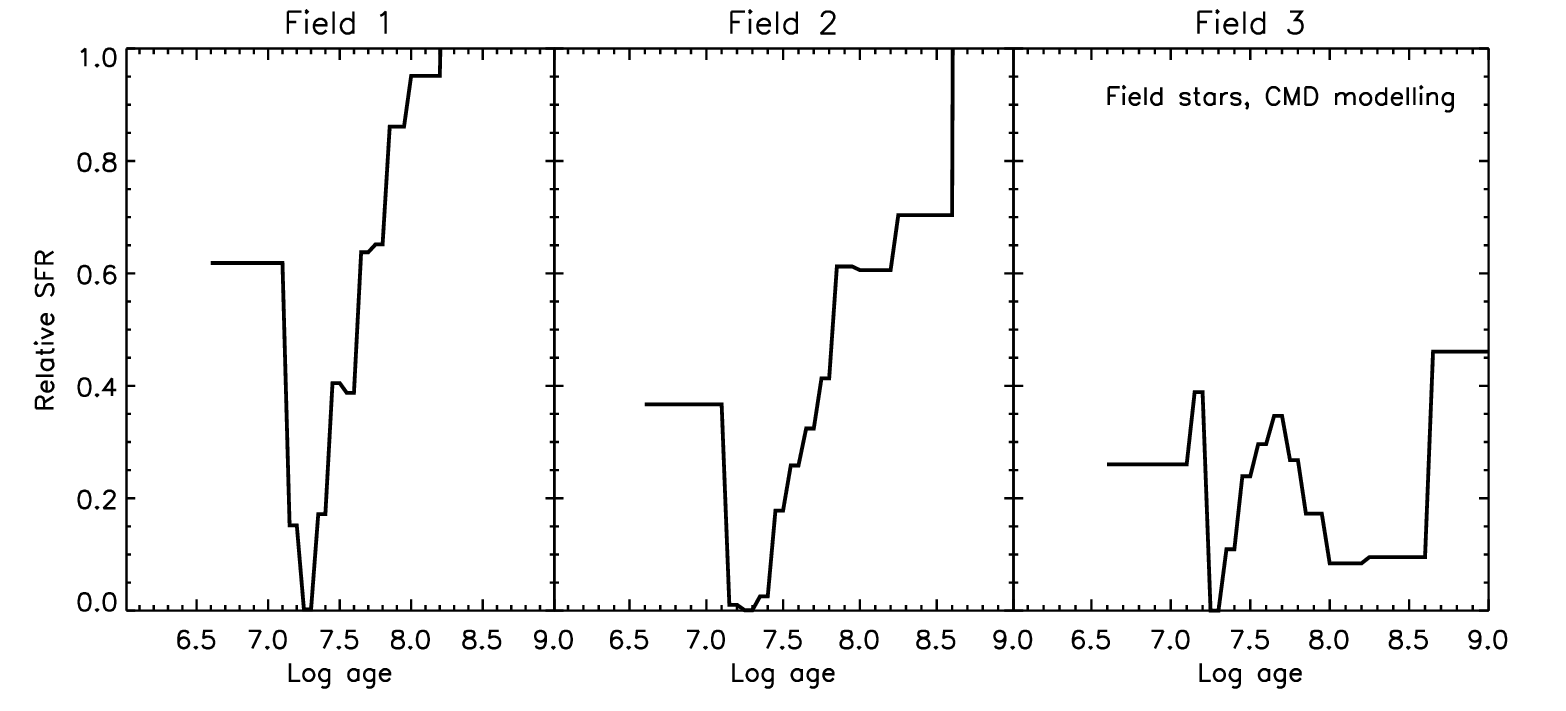} \\
  \noindent \includegraphics[width=14cm,height=4.5cm]{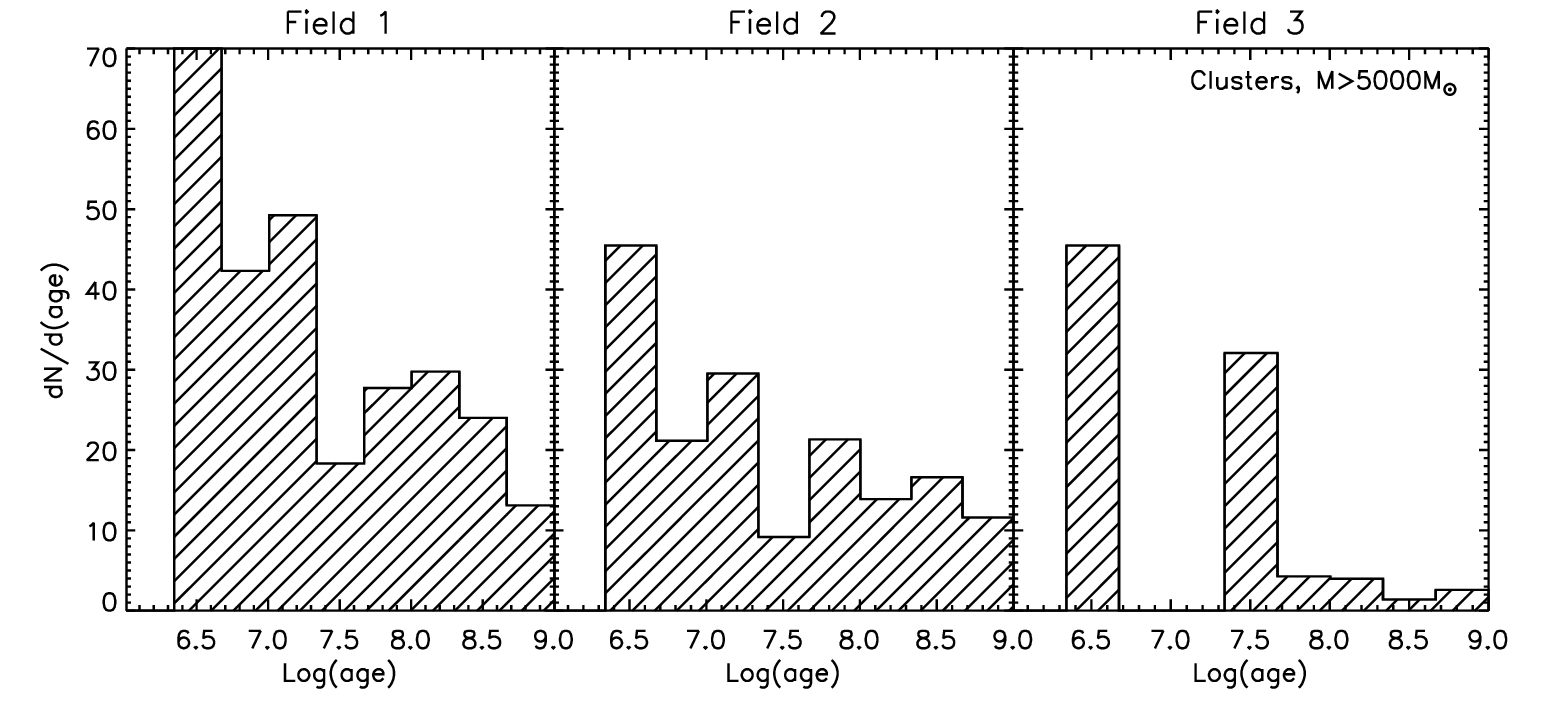} 
  \caption{\label{fig:claf123}Top: Hess diagrams for the three fields
   in NGC~1313. Centre: Field star formation histories obtained by
   modelling the CMDs. Bottom: Age distribution of clusters more
   massive than 5000 M$_\odot$.
  }
\end{figure}

The identification of star clusters remains challenging in star 
forming galaxies, and much less is known about star cluster populations in
spiral galaxies compared to the vast amount of data collected for
ellipticals and S0's over the past decade. It is not yet clear, for example, 
if there is a constant formation efficiency for (bound) clusters 
relative to ``field'' stars.  We have undertaken a survey of star
clusters in five nearby spirals, using the
\emph{Advanced Camera for Surveys} (ACS) on board the \emph{Hubble Space
Telescope} (HST).  Based on our ground-based imaging 
\cite[(Larsen \& Richtler 1999)]{lr99} the HST 
sample was designed to span a range from poor to rich cluster populations,
largely reflecting differences in the star formation rate.
The five galaxies are (roughly in order of richness): NGC~4395, NGC~45,
NGC~7793, NGC~1313 and NGC~5236. All are at distances of 4--5 Mpc, 
so that one ACS/WFC pixel corresponds to a linear scale of 
about 1 pc. With typical half-light radii of 2--4 pc, clusters
are thus readily identifiable as extended objects. In addition to ACS $BVI$
data, we include WFPC2 $U$-band imaging to allow
reliable age-dating of the clusters.  
First results for clusters in NGC~45, including the surprising detection
of a rich population of old GCs, have been presented in 
\cite{mora07}, and analysis of the
full sample of galaxies is currently in progress (Mora et al., in prep.).

  The galaxies in our sample are close enough to also resolve significant 
numbers of field stars, and an important aspect of our study is to link
the properties of field stars and cluster populations.  In the following 
we describe our first steps towards achieving this goal in the 
galaxy NGC~1313.

\section{Field Stars and Clusters in NGC~1313}

NGC~1313 is a nearby ($m-M=28.08$; M{\'e}ndez et al.\ 2002) late-type
barred spiral with a disturbed optical morphology, in which our ground-based
data revealed a rich population of massive young star clusters.
Our three HST/ACS pointings 
are illustrated in Figure~\ref{fig:n1313}: one includes the northern
spiral arm, the second includes the bar and the third pointing covers
a disturbed region of the galaxy, some $2.5^\prime$ ($\sim3$ kpc
projected) west of the main bar.

The ($V-I, M_V$) Hess diagrams for stars in the three fields are
shown in the top panel of Figure~\ref{fig:claf123}. Photometry was done
using the DAOPHOT package \cite[(Stetson 1987)]{stet87}; stars are detected
to $M_V\sim-2$ but the detection completeness drops rapidly
below $M_V\sim-3$.  Isochrones from the Padova group 
\cite[(Girardi et al.\ 2002)]{gir02} are shown for $Z=0.008$.
Already from a visual inspection 
of the Hess diagrams, one notes significant differences between 
the three fields: in particular, there is a relative lack of cool giants 
and/or supergiants
fainter than $M_V\sim-4$ in Field 3 and a corresponding excess of 
main sequence stars at $M_V\sim-3$, suggesting that star formation ramped 
up in this field some $10^8$ years ago.

By employing CMD modelling techniques we can obtain more detailed 
and quantitative information about the recent ($\sim$few times $10^8$ years) 
star formation histories in these fields. We used the Padova isochrones 
to construct model CMDs for
various star formation histories, including a realistic modelling of
photometric errors and completeness and (age-dependent) reddening. It is
a challenging test for the isochrones to match this region of the CMDs,
which includes the most massive stars, and we do not achieve perfect
fits. The best fits are obtained for $Z=0.008$ -- models of higher or
lower metallicities do not match the location of the red and blue core
He burning stars nearly as well.
In spite of significant uncertainties, a general picture of differences 
between the three fields can be constructed
(Fig.~\ref{fig:claf123}, centre): In Field 3, star formation indeed appears to
have ramped up about $10^8$ years ago, quite unlike the other two fields
where the star formation rate
has been roughly constant or decreasing over the last
several 100 Myrs. The dip to zero star formation at $\sim10^{7.2}$
years in all three fields is almost certainly due to difficulties with the
modelling of red supergiants - a similar
problem occurs around this age when ages are derived for star 
clusters from their integrated colours (e.g.\ Mora et al.\ 2007a).

We do not necessarily expect a 1:1 correspondence 
between field star and cluster age distributions, since the exact relation 
between the two may be affected by cluster disruption, 
variations in the cluster formation efficiency, mass functions etc.
Nevertheless, major bursts of star formation should be
visible in both the cluster- and field star age distributions. The bottom
panel of Fig.~\ref{fig:claf123} shows the age distributions of star
clusters with estimated masses $>5000$ M$_\odot$, derived by comparing their
integrated UBVI colours with SSP models (in this case $Z=0.008$
models taken from the website of the Padova group). Note that the histograms
show the number of clusters per \emph{linear} age interval, even 
though the age axis itself is logarithmic.
At least in a qualitative
sense, the cluster age distributions appear consistent with the field
star age distributions, with an excess of clusters around $\log t \approx 7.5$
in Field 3 (the bin at $\log t = 6.5$ only represents a single cluster). To 
give a more quantitative estimate of differences, we calculated
the number ratio of clusters older than $10^8$ years ($N_{\rm old}$) to
those younger than $10^8$ years ($N_{\rm young}$) for the three fields,
and obtain $N_{\rm old}/N_{\rm young}$ = 4.9 (Field 1), 6.1 (Field 2)
and 2.2 (Field 3). This again seems consistent with a relative increase in 
star formation activity in Field 3 in the last $10^8$ years.

\section{Concluding remarks}

With ACS imaging, we are now in a position to simultaneously constrain the 
star and cluster age distributions for galaxies outside the Local Group. 
Our first results provide evidence for
an increase in star formation rate $\sim10^8$ years ago in one of 
three fields in NGC~1313,
which is reflected in \emph{both} the field star and cluster age 
distributions.

\begin{acknowledgments}
SSL is grateful to the Leids Kerkhoven-Bosscha Fonds for supporting
attendance of the Symposium.
\end{acknowledgments}

\begin{discussion}

\discuss{B{\"o}ker}{Is there any evidence for multiple stellar populations
  in any of the (young) massive star clusters? This would argue for them
  being stripped nuclei.}

\discuss{Larsen}{So far no such evidence has emerged, although it is also
  difficult to rule out. However, most of these clusters do seem to be part 
  of the normal hierarchy of star formation in their parent galaxies, e.g.\
  the youngest ones tend to be found in/near spiral arms.}

\discuss{Meixner}{Are you planning to get near-IR photometry to complement
  your dataset?}

\discuss{Larsen}{Yes, absolutely. In fact the Wide Field Camera 3 is likely
  to reinvigorate this whole field thanks to its panchromatic wavelength
  coverage from the UV to the near-IR.}

\discuss{Fritze}{I am not convinced that the LFs for \emph{all} YSC systems
  are power laws as you suggested in one of your first slides. Peter Anders
  recently has done a very careful re-analysis of WFPC2 data for the
  Antennae and found clear and solid evidence for a turn-over in the YSC LF.
  This result casts serious doubts on the universality of the SF and SCF
  process. Apparently the enhanced ambient pressure in a major merger can act
  to preferentially produce massive clusters.
  }

\discuss{Larsen}{I am very curious to learn more about these results.
  However, I would note that new ACS data are now available for the
  Antennae so I suspect that the final word has not yet been said.
}

\end{discussion}

\end{document}